\documentclass{PoS}

\usepackage{cite}

\newcommand{\bra}{\left\langle}
\newcommand{\ket}{\right\rangle}
\newcommand{\bbra}{\langle\!\langle}
\newcommand{\kket}{\rangle\!\rangle}

\renewcommand{\det}{\mbox{det}\:}
\newcommand{\tr}{ {\rm tr} \: }

\newcommand{\half}{\frac{1}{2}}

\title{Mesonic correlators at non-zero baryon chemical potential}

\ShortTitle{Mesonic correlators at non-zero baryon chemical potential}

\author{\speaker{Aleksandr Nikolaev}, Gert Aarts, Chris Allton, Davide De Boni, Jonas Glesaaen, Simon Hands \\
        Department of Physics, Swansea University, Swansea SA2 8PP, United Kingdom\\
        E-mail: \email{\{aleksandr.nikolaev, g.aarts, c.allton, s.hands\}@swansea.ac.uk}, \email{dade89.10@gmail.com}, \email{jonas@glesaaen.com}
        }

\author{Benjamin J\"ager\\
        CP$^3$-Origins \& Danish IAS, Department of Mathematics and Computer Science, University of Southern Denmark, 5230 Odense M, Denmark\\
        E-mail: \email{jaeger@cp3.sdu.dk}}

\author{Jon-Ivar Skullerud\\
        Dept.\ of Theoretical Physics, National University of Ireland Maynooth, County Kildare, Ireland\\
        E-mail: \email{jonivar@thphys.nuim.ie}}

\author{Liang-Kai Wu\\
        Faculty of Science, Jiangsu University, Zhenjiang, 212013 \& Key Laboratory of Quark and Lepton Physics (MOE), Central China Normal University, Wuhan 430079, China\\
        E-mail: \email{wuliangkai@ujs.edu.cn}}

\abstract{In order to study the fate of mesons in thermal QCD at finite baryon chemical potential, we consider light mesonic correlation functions using the Taylor expansion to ${\cal O}((\mu/T)^2)$, in both the hadronic and quark-gluon plasma phases. We use the FASTSUM anisotropic fixed-scale lattices with $N_f = 2+1$ flavours of Wilson fermion. We find that mesonic correlators are sensitive to finite-density corrections and that the second-order terms indicate the chiral crossover in the vector and axial-vector channels.}

\FullConference{37th International Symposium on Lattice Field Theory - Lattice2019\\
		16-22 June 2019\\
		Wuhan, China}

\begin{document}

\section{Introduction}

The behaviour of light hadrons at nonzero temperature and baryon density is an important open question in QCD and is highly relevant for the experimental heavy-ion programme and the structure of neutron stars. Lattice QCD is directly applicable at nonzero temperature but vanishing density -- see e.g.\ Refs.\ \cite{Aarts:2017rrl,Aarts:2018glk} for a study of hyperons at finite temperature -- however at nonzero baryon chemical potential the sign problem \cite{Aarts:2015tyj} prohibits a direct application of standard Monte Carlo methods. One of the many indirect approaches available is to expand observables in a Taylor series in $\mu/T$ and compute the Taylor coefficients at $\mu=0$. This may give quantitative results for small $\mu/T$, provided the expansion converges and the higher-order terms, which are typically very noisy, can be evaluated precisely. This approach has been applied with great success to properties of the thermal crossover and bulk thermodynamic quantities, see e.g.\ the classic papers \cite{Allton:2002zi,Allton:2005gk,Gavai:2003mf} and the review \cite{Ding:2017giu}.
Here we are interested in hadrons at small but nonzero chemical potential, for which much less is known, with the notable exception of the pioneering paper \cite{Choe:2001cq}.

\section{Taylor expansion of mesonic correlators} 

We consider mesonic correlators of the form
\begin{equation}
  G_H(x) = \left\langle J_H(x) J_H^\dagger(0) \right\rangle,
\end{equation}
where $J_H = \bar\psi \Gamma_H \psi$ is a meson operator in the isotriplet channel $H$, with $\Gamma_H$ the appropriate combination of $\gamma$ matrices (we follow the notation of Ref.\ \cite{Aarts:2005hg}), and the brackets indicate a thermal average. Where possible, we drop the $H$ subscript below. After integrating out the quark fields, the correlator can be written as
\begin{equation}
    G(x)  = \frac{\bbra g(x) \det M\kket}{\bbra \det M \kket} \equiv   \bra g(x)\ket,    
\end{equation}
where the double brackets indicate the expectation value with respect to the gluonic fields only, $\det M$ is the determinant of the fermion matrix $M$, and 
\begin{equation}
    g(x)  = \tr \Bigl[ S(x)\Gamma S(-x)\Gamma^\dagger \Bigr]
\end{equation}
are the contracted quark propagators. Dependence on the chemical potential arises from both the fermion determinant and the quark propagators. We consider QCD with $N_f=2+1$ flavours with degenerate $u$ and $d$ quarks. Chemical potentials are assumed as $\mu_u = \mu_d \equiv \mu$ ($\mu$ here and below denotes the quark chemical potential) and $\mu_s = 0$, which partially simplifies the expressions. 
 
 Following Ref.\ \cite{Choe:2001cq}, we expand the correlator to ${\cal O}\left((\mu/T)^2\right)$, i.e.,
 \begin{equation}
\label{eq:expansion_start}
  G(x) = G(x)\Big|_{\mu = 0} + \frac{\mu}{T} T G^\prime(x)\Big|_{\mu = 0} + \frac{1}{2} \left(\frac{\mu}{T} \right)^2 T^2 G^{\prime\prime}(x)\Big|_{\mu = 0} + {\cal O}\left( \frac{\mu^4}{T^4} \right),
\end{equation}
where derivatives with respect to $\mu$ are indicated with a prime ($\prime$).
At the first order we find 
\begin{equation}
\label{eq:first_term}
G^\prime(x) =
    \bra g^{\prime}(x)\ket
    + \bra g(x)\frac{\det M^\prime}{\det M} \ket
    - \bra \frac{\det M^\prime}{\det M}\ket \bra g(x)\ket,
\end{equation}
but it is easy to see that this contribution vanishes when evaluated at $\mu=0$, as is the case for the
${\cal O}\left( \mu^3 / T^3 \right)$ contribution. At second order, a number of terms appear, which can be organised as
\begin{eqnarray}
G^{\prime\prime}(x) &=& \bra g^{\prime\prime}(x)\ket 
+2 \bra \frac{\det M^\prime}{\det M} g^\prime(x)\ket 
-2 \bra \frac{\det M^\prime}{\det M}\ket \bra g^\prime(x)\ket \nonumber \\ &&
+
\bra \frac{\det M^{\prime\prime}}{\det M} g(x)\ket - \bra \frac{\det M^{\prime\prime}}{\det M} \ket\bra g(x)\ket \label{eq:G_primeprime}
\\ && 
-2 \left( \bra \frac{\det M^\prime}{\det M} g(x)\ket   
- \bra \frac{\det M^\prime}{\det M}\ket\bra g(x)\ket\right) 
\bra \frac{\det M^\prime}{\det M}\ket\,. \nonumber
\end{eqnarray}
Here the first term $\bra g^{\prime\prime} \ket$ is the connected contribution, while all the remaining combinations contain disconnected contributions, which vanish at asymptotically high temperature, due to asymptotic freedom. When evaluating $G^{\prime\prime}(x)$ at $\mu=0$, we note that the terms with $\bra \det M^{\prime}/\det M\ket=\bra \tr M^{-1}M^{\prime}\ket$ vanish, since this quantity is proportional to the quark number density, which is identically zero at vanishing chemical potential.

In order to investigate what to expect at high temperature, we have evaluated the leading term in perturbation theory, which amounts to the computation of $\bra g^{\prime\prime}(x)\ket$ for free quarks, extending the setup in Ref.\ \cite{Aarts:2005hg} to non-zero chemical potential. We consider vanishing external momentum.  In the continuum and for massless quarks, the $\mu^2$-correction may be written as 
\begin{equation}
\label{eq:free_cont_Gprpr}
 G^{\prime \prime}_H(\tau)/T\Big|_{\mu=0} =
 \frac{N_c}{\pi^2} \left[
 a_H^{(1)}+a_H^{(2)}-\frac{1}{12} \left( a_H^{(1)}-a_H^{(2)}\right) h(u) \right],
\end{equation}
where the coefficients $a_H^{(i)}$ depend on the channel \cite{Aarts:2005hg} and
\begin{eqnarray}
\label{eq:h}
h(u) = \frac{ 3u (\pi^2 - u^2-2) + u(\pi^2 - u^2 + 6)\cos(2u) - 2 (\pi^2 - 3u^2)\sin(2u)}{\sin^3(u)},
\end{eqnarray}
with $u=2\pi T(\tau -1/2T)$, $-\pi<u<\pi$. Note that $\tau$-dependence is contained in $h(u)$ only and is hence identical in all channels. We have also extended the calculation to free lattice fermions \cite{Aarts:2005hg}.

\begin{figure}[t]
\begin{center}
  \includegraphics[width=0.49\linewidth]{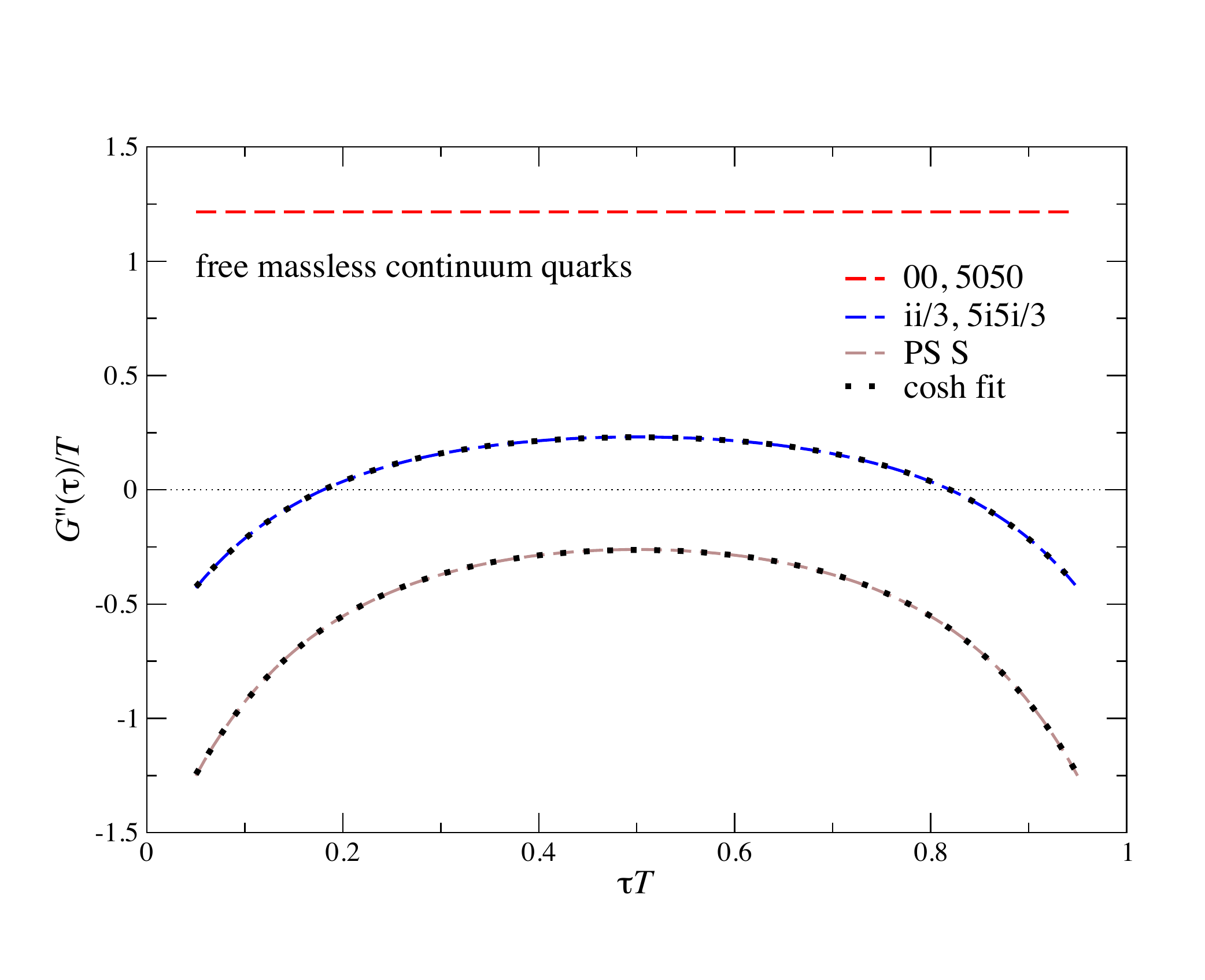}
 \includegraphics[width=0.49\linewidth]{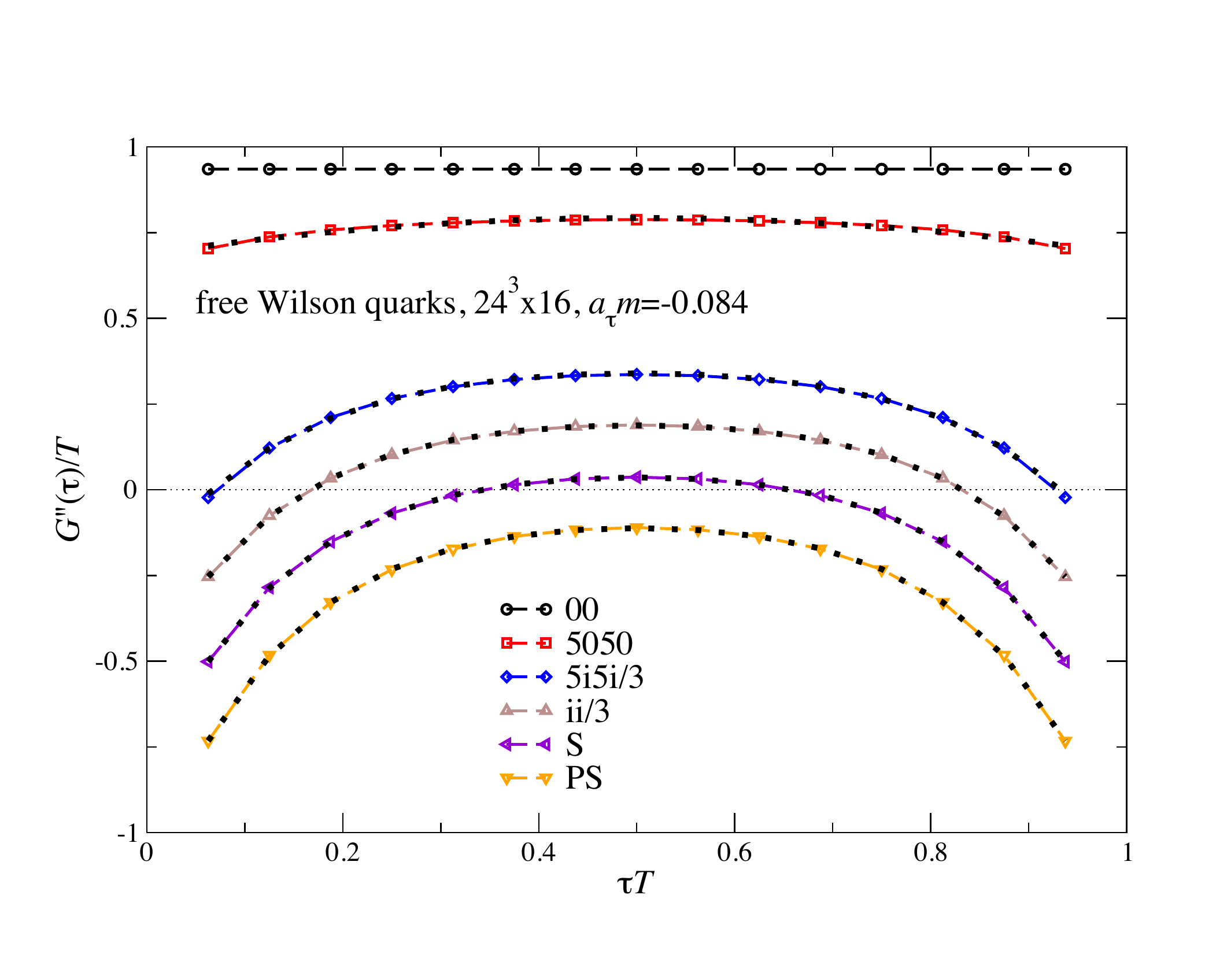}
  \caption{Second-order correction $G^{\prime\prime}(\tau)/T$ for free massless quarks in the continuum (left) and  free Wilson quarks on a $24^3\times 16$ lattice (right).
  The dotted lines show the result from the fit (\ref{eq:one_cosh}), with 
  $c_2 \sim 7$ (continuum) and $c_2 \sim 7.4$ (lattice). Here ``00'' and ``5050'' mean temporal components of the vector and axial vector currents respectively, ``ii/3'' and ``5i5i/3'' indicate averaged spatial components of the vector and axial vector currents, and ``PS'' and ``S'' the pseudoscalar and scalar correlators.}
  \label{fig:free}
\end{center}
\end{figure}

The resulting contributions are shown in Fig.~\ref{fig:free} for continuum (left) and Wilson (right) fermions on a $24^3\times 16$ lattice.
The channels are denoted by the $\gamma$ matrices appearing in the meson operator: 
  $\Gamma_H=\gamma_0$ (00, density), $\gamma_5\gamma_0$ (5050, axial density),
  $\gamma_i$ (ii/3, vector), $\gamma_5\gamma_i$ (5i5i/3, axial-vector),
  $\gamma_5$ (PS, pseudoscalar), $1\!\!1$ (S, scalar). The (axial) densities are conserved in the free theory and hence these correlators are independent of Euclidean time. The observed degeneracy for massless continuum quarks (left) is absent for Wilson fermions (right). Perhaps surprisingly, the second-order corrections are well described by the following inverted ($c_1<0$) one-cosh fit,
\begin{equation}
\label{eq:one_cosh}
G''(\tau)/T = c_0 + c_1 \cosh[c_2(\tau T - 1/2)],
\end{equation}
with $c_2\sim 7$ (continuum) and $c_2\sim 7.4$ (lattice), even though the exact $\tau$-dependence in the continuum is given by Eq.\ (\ref{eq:h}). These fits are shown on the Fig.~\ref{fig:free} with dotted lines.

Finally, we note that since the corrections do not have a definite sign in all channels, the inclusion of the ${\cal O}((\mu/T)^2)$ correction has a non-monotonic effect as a function of $\tau$ on the full correlator, when the first two terms in the Taylor expansion are combined.

\section{Vector and axial-vector channels}

We have evaluated the second-order correction to meson correlators on our FASTSUM~\cite{fastsum} Generation 2 ensembles. These are fixed-scale, anisotropic lattices, with an anisotropy of $a_s/a_\tau= 3.5$, $N_f=2+1$ Wilson-clover fermions, a pion mass of $m_\pi=384(4)$ MeV, and a physical strange quark. 
Tuning and the ensemble at the lowest temperatures have
been provided by HadSpec~\cite{Edwards:2008ja}.
Details of the ensembles are given in Table \ref{tab2} and a further discussion of the FASTSUM ensembles can be found in Refs.\ \cite{Aarts:2019hrg,Aarts:2014cda,Aarts:2014nba}.  We employ \cite{openqcd-fastsum} a modification of openQCD \cite{openqcd}, to include anisotropy and stout-smearing, supplemented with a stand-alone spectroscopy code \cite{openqcd-hadspec}.

\begin{table}[b]
\begin{center}
\begin{tabular}{|c||cccc | cccc|}
\hline
 $N_\tau$ & 128$^*$ & 40 & 36 & 32 & 28 & 24 & 20 & 16 \\
\hline
$T$ [MeV]  	  & 44	 & 141  & 156   & 176 & 201  & 235  & 281 & 352 \\
$T/T_c$	      & 0.24 & 0.76  & 0.84   & 0.95 & 1.09  & 1.27  & 1.52 & 1.90 \\
$N_{\rm cfg}$ & 139	 & 501 & 501 & 1000  & 1001 & 1001 & 1000 & 1001\\
\hline
\end{tabular}
\caption{Generation 2 ensembles, $m_\pi=384(4)$ MeV, lattice size $24^3 \times N_\tau$, spatial lattice spacing $a_s=0.1227(8)$ fm, temporal lattice spacing $a_\tau^{-1}=5.63(4)$ GeV, anisotropy $a_s/a_\tau=3.5$. The choice of parameters and the ensemble at the lowest temperature are courtesy of the HadSpec collaboration \cite{Edwards:2008ja}.}
\label{tab2}
\end{center}
\end{table}

\begin{figure}[t]
\begin{center}
  \includegraphics[width=0.49\linewidth]{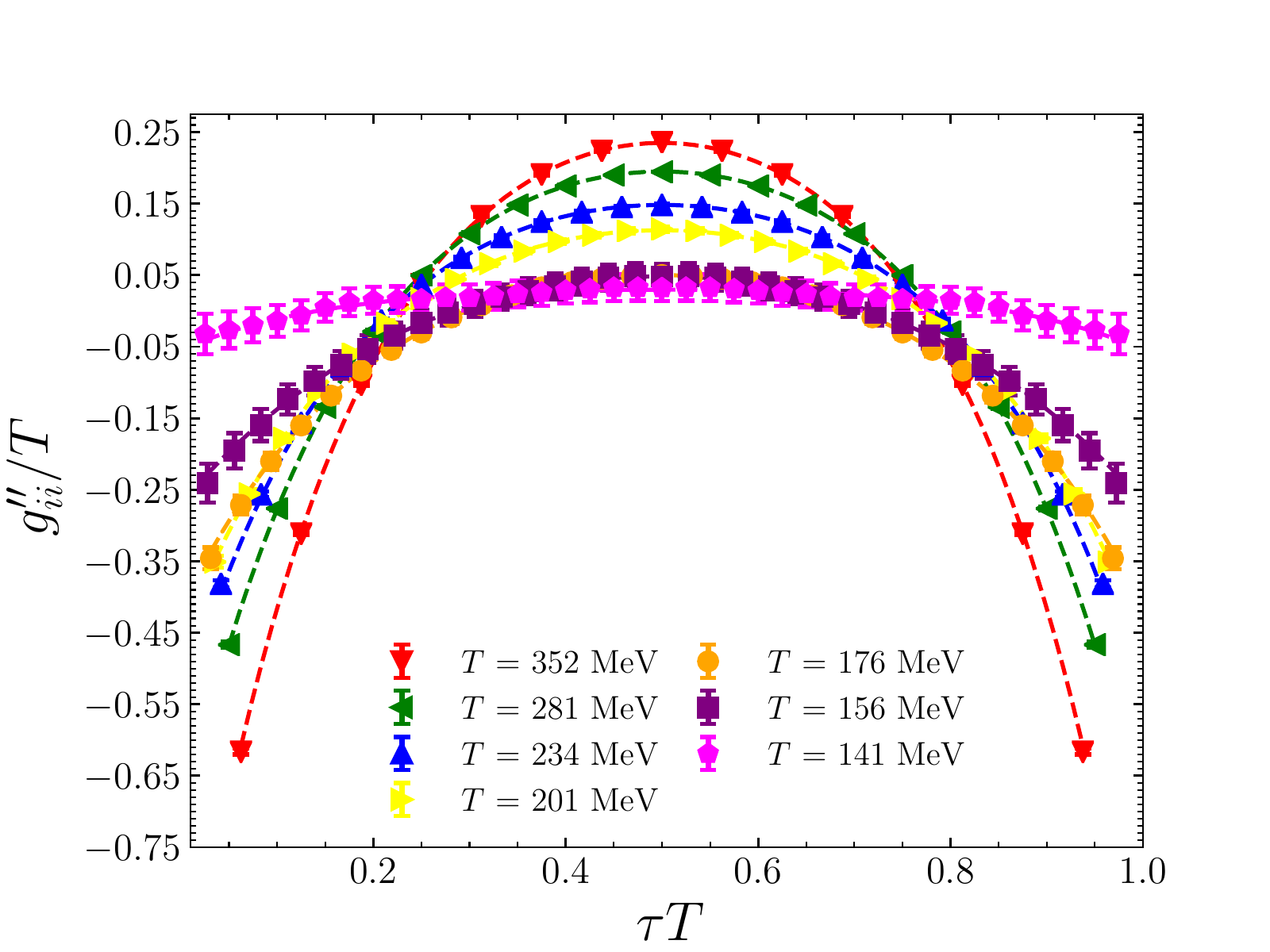}
  \includegraphics[width=0.49\linewidth]{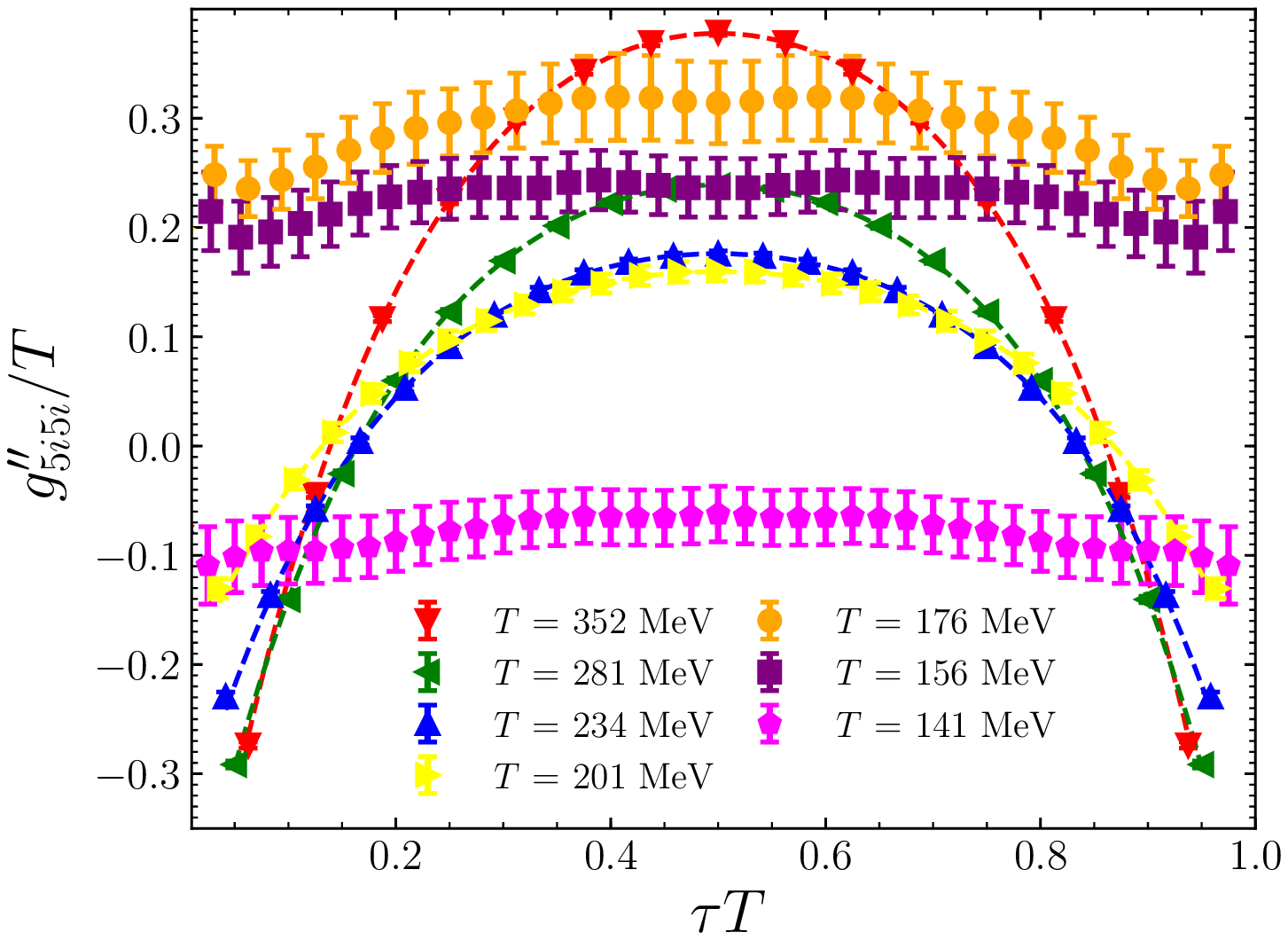}
  \caption{Connected contribution $\bra g''(\tau)\ket/T$ in the vector (left) and axial-vector (right) channel. The dashed lines are fits according to Eq.\ (\ref{eq:one_cosh}).}
  \label{fig:g}
\end{center}
\end{figure}
\vspace*{0.4cm}
\begin{figure}[t]
\begin{center}
  \includegraphics[width=0.49\linewidth]{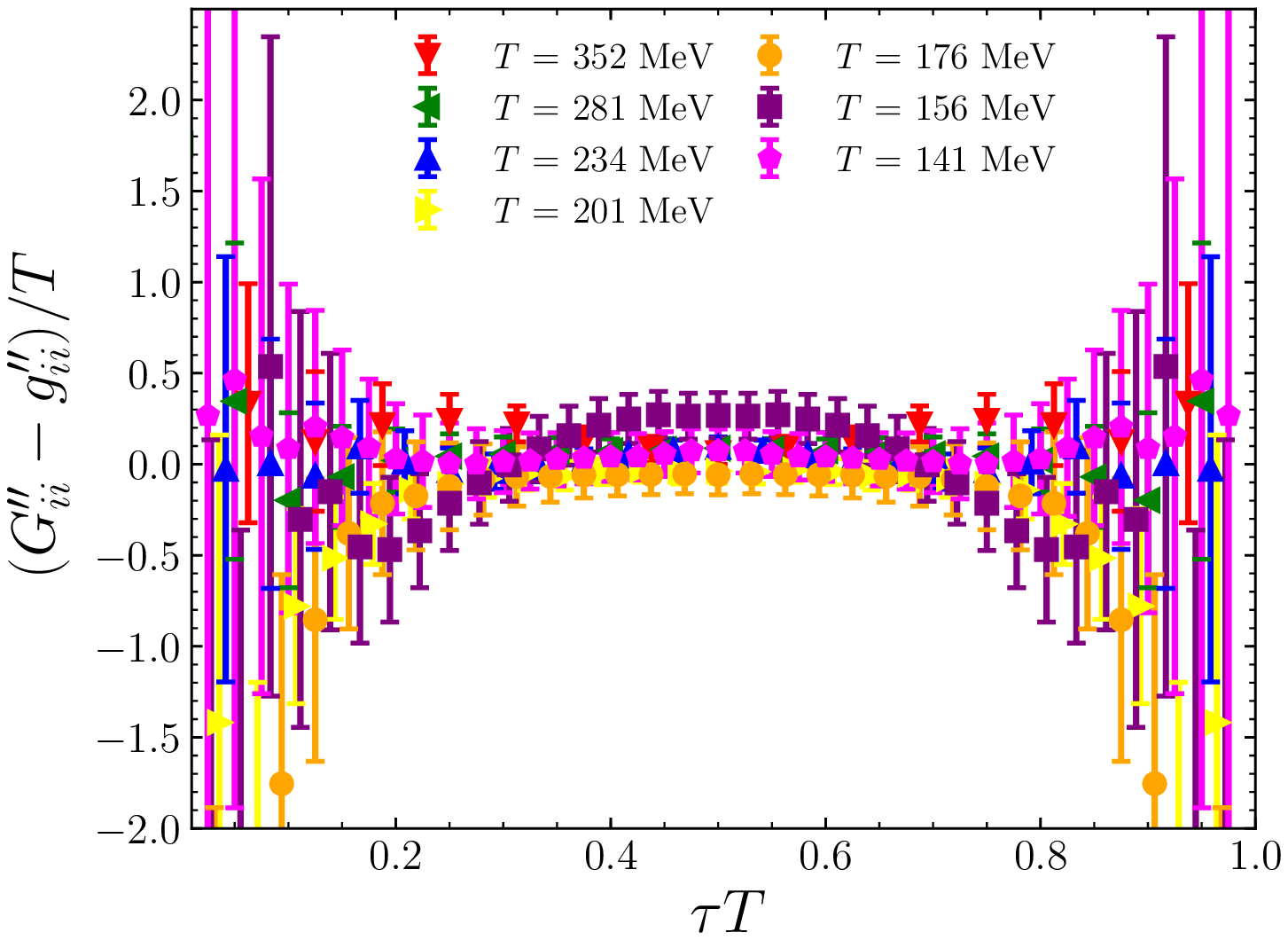}
   \includegraphics[width=0.49\linewidth]{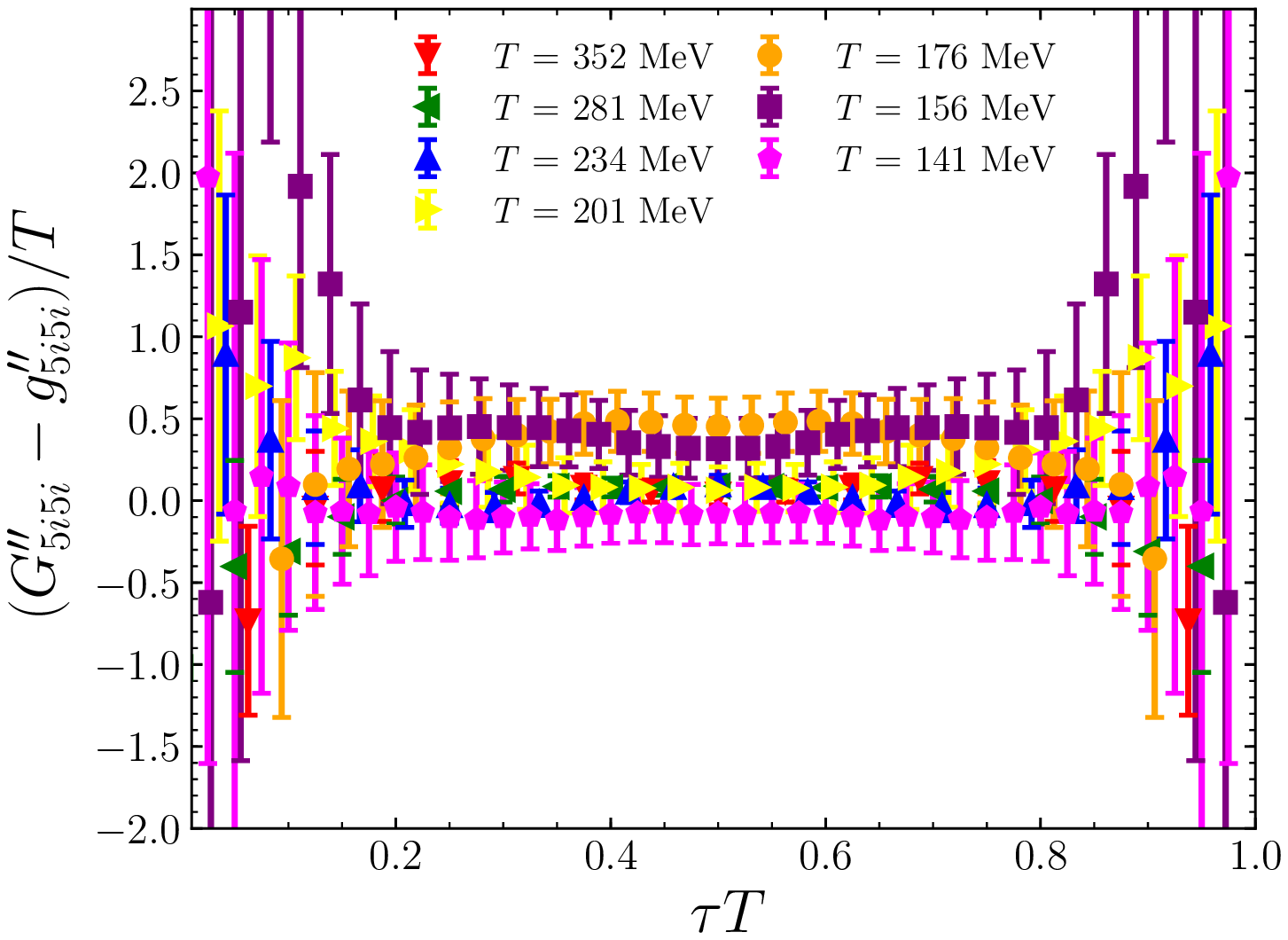}
  \caption{Disconnected contribution $[ G''(\tau) - \bra  g''(\tau) \ket]/T$ in the vector (left) and axial-vector (right) channel.}
    \label{fig:dis}
\end{center}
\end{figure}

We have computed the second-order correction~(\ref{eq:G_primeprime}) in a variety of channels, here we present results in the vector and axial-vector channels. 
We start by discussing the connected contribution $\bra g^{\prime\prime}(\tau)\ket$ in Fig.~\ref{fig:g}, which can be computed with reasonable accuracy and is non-zero at all temperatures. At high temperatures, we note a similarity with the results obtained in the free theory (see Fig.~\ref{fig:free}), which hence provide a useful benchmark. We find that the data at the higher temperatures can again be described by the ansatz (\ref{eq:one_cosh}), but with a slightly lower coefficient $c_2$: 5.77(25) for the vector and 5.62(48) for the axial-vector channel. These fits are shown by the dashed lines in Fig.~\ref{fig:g}. The transition between the confined and deconfined phase is a crossover, with an estimation for the pseudocritical temperature $T_{\rm pc}$ given by $181(1)$ MeV, when determined via the renormalised chiral condensate \cite{Aarts:2019hrg}, and $185(4)$, when the renormalised Polyakov loop is used \cite{Aarts:2014nba}. We observe that this transition is also visible in the connected second-order contribution, where at the lower temperatures the data become nearly independent of $\tau$, especially deep in the hadronic phase. We observe that the transition is most pronounced in the axial-vector channel.

Next we show the disconnected contributions in the vector and axial-vector channels in Fig.~\ref{fig:dis}. These contributions are substantially noisier than the connected ones, and fluctuate close to zero. Interestingly, the noise dominates at short times, rather than around $\tau T=\half$. The disconnected contributions are expected to vanish at very high temperature, due to asymptotic freedom, and at very low temperature, where $\mu$-dependence only arises due to interactions with baryons, which is suppressed by the large baryon masses. Indeed, the indication of a non-zero signal is most prominent just below $T_{\rm pc}$, and at high temperatures these contributions are zero within the errorbars. 

\begin{figure}[t]
\begin{center}
  \includegraphics[width=0.49\linewidth]{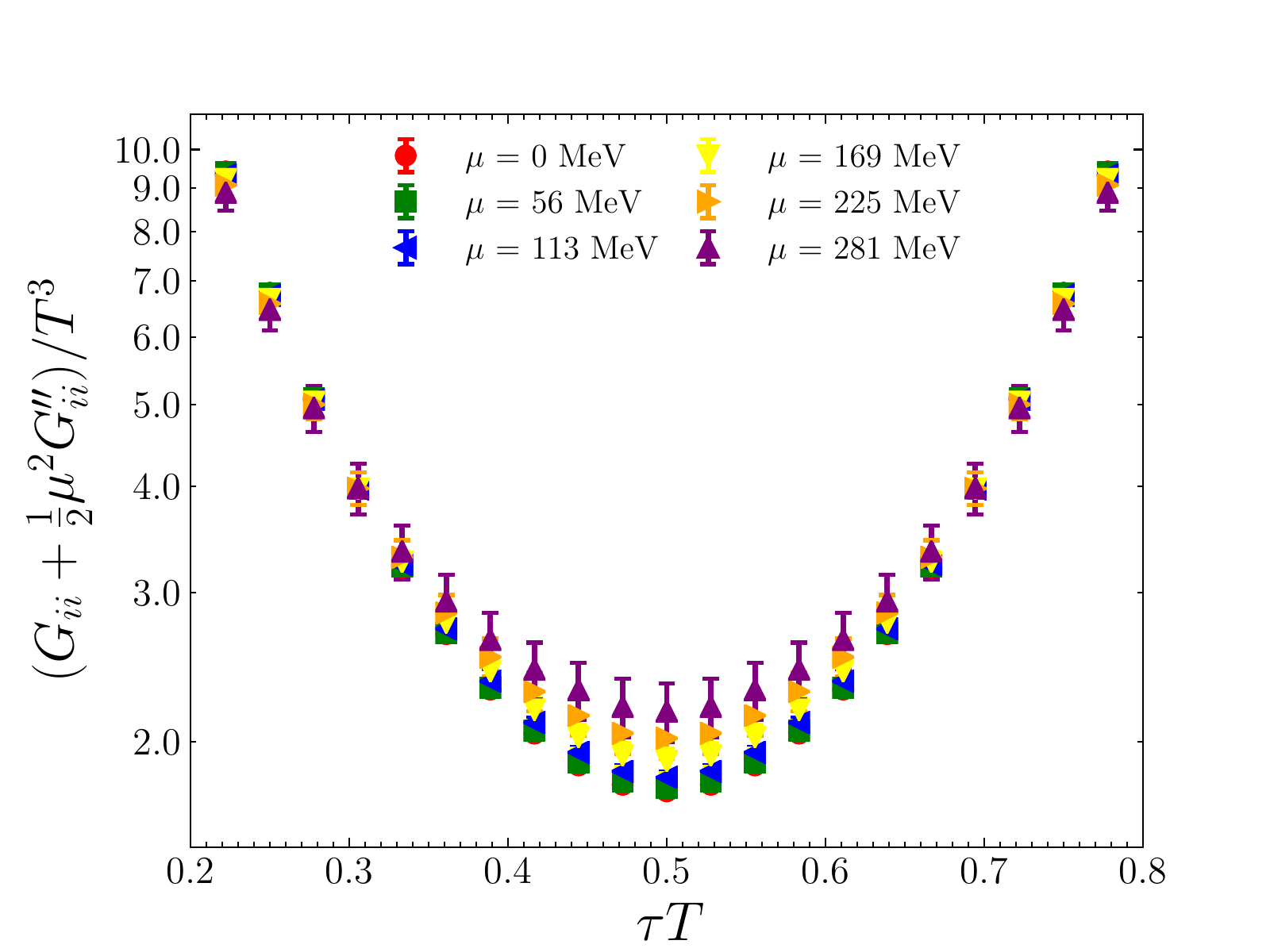}
  \includegraphics[width=0.49\linewidth]{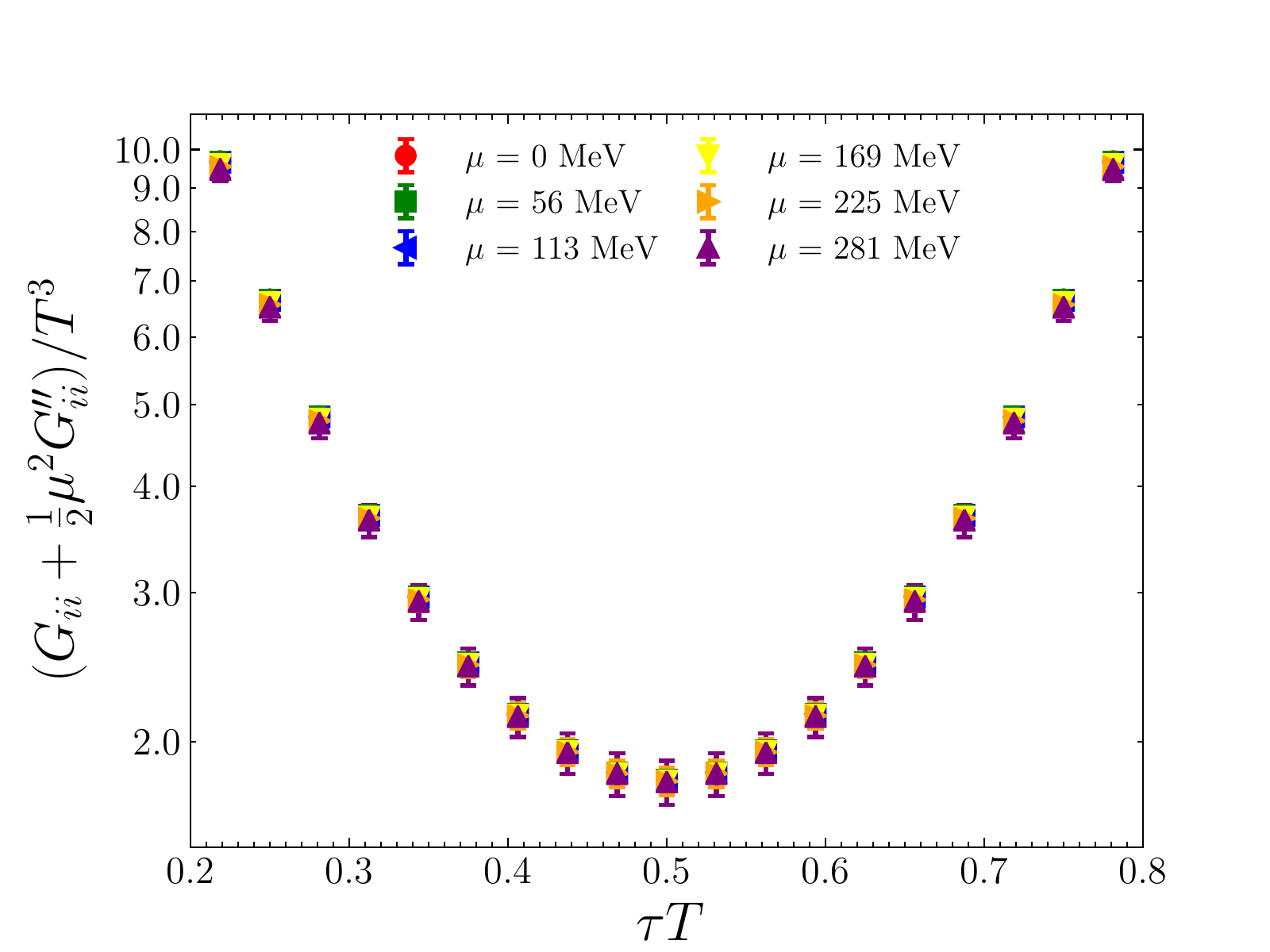}
  \caption{Vector ($\rho$-meson) correlator up to second-order in $\mu/T$ for various values of $0\leq \mu \leq 281$ MeV, at $T=156$ (left) and $176$ MeV (right).}
  \label{fig:Grho}
\end{center}
\end{figure}

Finally, in Fig.~\ref{fig:Grho} we show the complete correlator in the vector ($\rho$-meson) channel, combining the ${\cal O}(\mu^0)$ and all ${\cal O}(\mu^2)$ contributions, for six values of $\mu$ between $0$ and $281$ MeV at two temperatures in the hadronic phase (recall that $\mu$ is the quark chemical potential). We observe a reasonable signal, which is controlled by  the smallness of $\mu/T$. We also note a stronger $\mu$ dependence at the lower temperature.
Further analysis will allow us to estimate e.g.\ a $\mu$-dependent mass parameter and other spectral features, which yields the possibility to compare with effective model approaches, see e.g.\ Ref.\ \cite{Jung:2016yxl}.

\section{Conclusions}

We have investigated correlators of light mesons, employing a Taylor expansion to ${\cal O}((\mu/T)^2)$, where $\mu$ is the light quark chemical potential, using the FASTSUM anisotropic thermal ensembles in the hadronic phase and the quark-gluon plasma.
Focusing on the vector and axial-vector channels, we found qualitative agreement with the non-interacting theory at high temperature. The second-order correction in the axial-vector channel is especially sensitive to the confinement-deconfinement transition.
While in the deconfined phase the disconnected contributions are close to zero, in the hadronic phase they are very noisy in most mesonic channels. Increasing the number of noise vectors in the stochastic estimator does not appear to beneficial, and hence noise reduction techniques for disconnected diagrams are needed to make further progress.
Nevertheless, a reasonable signal-to-noise ratio is obtained for the vector channel in the hadronic phase, allowing for further studies of the role of a non-zero baryon density on the behaviour of mesons.

\acknowledgments

We are grateful for support from STFC via grants  ST/L000369/1 and ST/P00055X/1, to RFBR via grant 18-02-40126 mega, the Swansea Academy for Advanced Computing (SA$^2$C), SNF, ICHEC, COST Action CA15213 THOR, the European Research Council (ERC) under the European Union's Horizon 2020 research and innovation programme under grant agreement No 813942, and the Key Laboratory of Ministry of Education of China under Grant No.\ QLPL2018P01. Computing resources were made available by HPC Wales and Supercomputing Wales and we acknowledge PRACE for access to the Marconi-KNL system hosted by CINECA, Italy. This work used the DiRAC Extreme Scaling service and the DiRAC Blue Gene Q Shared Petaflop system at the University of Edinburgh, operated by the Edinburgh Parallel Computing Centre on behalf of the STFC DiRAC HPC Facility. This equipment was funded by BIS National E-infrastructure capital grant ST/K000411/1, STFC capital grants ST/H008845/1 and ST/R00238X/1, and STFC DiRAC Operations grants ST/K005804/1, ST/K005790/1 and ST/R001006/1. DiRAC is part of the National e-Infrastructure.

\end{document}